\documentclass[10pt,5p,twosided]{elsarticle}

\biboptions{comma,square,sort}

\usepackage[latin1]{inputenc}
\usepackage{etex}
\usepackage{amsmath}
\usepackage{amsfonts}
\usepackage{amssymb}
\usepackage{graphicx}
\usepackage{tikz}
\usepackage{color,soul}
\usepackage[all,cmtip]{xy}
\usepackage{dcolumn}
\usepackage[font=small,labelfont=bf, figurename=Fig.]{caption}
\usepackage{makeidx}
\usepackage{nomencl}

\makenomenclature

\usepackage{hyperref}
\usepackage{cleveref}
\hypersetup{colorlinks=true,linkcolor=blue,citecolor=blue}

\author{
	David P. Chassin$^{1,2}$, Sahand Behboodi$^2$ and Ned Djilali$^2$ 
\\
	$^1$\textit{SLAC National Accelerator Laboratory, Menlo Park CA (USA)}
\\
	$^2$\textit{Inst. for Integrated Energy Systems, and Mechanical Engineering Dept, University of Victoria, Victoria BC (Canada)}
}

\title{Optimal Subhourly Electricity Resource Dispatch Under Multiple Price Signals With High Renewable Generation Availability}

\journal{Applied Energy}

\newcommand{\reffig}[1]{Fig.~\ref{fig:#1}}

\newcommand{\dd}[2]{\frac{d#1}{d#2}}

\newcommand{\pd}[2]{\frac{\partial #1}{\partial #2}}
\newcommand{\pdn}[3]{\frac{\partial ^{#1}#2}{\partial #3^{#1}}}
\newcommand{\eq}[1]{\begin{align*}#1\end{align*}}
\newcommand{\eqn}[2]{\begin{align}\label {eq:#1}#2\end{align}}
\renewcommand{\exp}[1]{e^{#1}}

\newcommand{\Matrix}[1]{\begin{bmatrix}#1\end{bmatrix}}

\begin{document}

\begin{abstract}

This paper proposes a system-wide optimal resource dispatch strategy that enables a shift from a primarily energy cost-based approach, to a strategy using simultaneous price signals for energy, power and ramping behavior. A formal method to compute the optimal sub-hourly power trajectory is derived for a system when the price of energy and ramping are both significant. Optimal control functions are obtained in both time and frequency domains, and a discrete-time solution suitable for periodic feedback control systems is presented. The method is applied to North America Western Interconnection  for the planning year 2024, and it is shown that an optimal dispatch strategy that simultaneously considers both the cost of energy and the cost of ramping  leads to significant cost savings in systems with high levels of renewable generation: the savings exceed 25\% of the total system operating cost for a 50\% renewables scenario.

\end{abstract}

\begin{keyword}
Electricity pricing \sep
bulk electric system \sep
optimal energy dispatch \sep
optimal ramping \sep
renewable integration \sep
resource allocation
\end{keyword}

\maketitle

\section*{Highlights}

\begin{itemize}

\item A method to minimize the cost of subhourly dispatch of bulk electric power systems.

\item Dispatch based on simultaneous use of energy and ramping costs yields significant savings 

\item Savings from optimal dispatch increase as transmission constraints increase. 

\item Savings from optimal dispatch increase as variable generation increases.

\end{itemize}

\nomenclature{$E_T$}{Energy over $T$ in MWh.}
\nomenclature{$Q_E$}{Scheduled load in MW.}
\nomenclature{$Q_T$}{Terminal load in MW.}
\nomenclature{$k$}{Discrete time step in p.u. $t_s$.}
\nomenclature{$Q^*$}{Discrete power in MW.}
\nomenclature{$\dot Q^*$}{Discrete power at next time step in MW.}
\nomenclature{$C_{base}$}{Cost associated with base case control in \$.}
\nomenclature{$C^*$}{Cost associated with discrete time control in \$.}
\nomenclature{$E(t)$}{Energy over the time interval 0 to $t$ in MWh.}
\nomenclature{$T$}{Interval terminating time in hours.}
\nomenclature{$Q(t)$}{Power in MW.}
\nomenclature{$P(Q)$}{Power price function in \$/MWh.}
\nomenclature{$a$}{Marginal price of energy in \$/MW$^2\cdot$h.}
\nomenclature{$b$}{Marginal price of power in \$/MW$^2$.}
\nomenclature{$c$}{Marginal price of ramping in \$$\cdot$h/MW$^2$.}
\nomenclature{$\dot Q$}{Ramping in MW/h.}
\nomenclature{$R(Q,\dot Q)$}{Ramping price function in \$/MW.}
\nomenclature{$t$}{Time domain real variable in hours.}
\nomenclature{$C(t)$}{Cost over the time interval 0 to $t$ in \$.}
\nomenclature{$t_s$}{Time step in seconds.}
\nomenclature{$\mu$}{Lagrange multiplier (including $Q_z$) in \$/MWh.}
\nomenclature{$\lambda$}{Lagrange multiplier (excluding $Q_z$)  in \$/MWh.}
\nomenclature{$G(t,Q,\dot Q)$}{Cost Lagrangian in \$.}
\nomenclature{$\ddot Q$}{Ramping rate of change in MW/h$^2$.}
\nomenclature{$s$}{Frequency domain complex variable in h$^{-1}$.}
\nomenclature{$\omega$}{Square root of energy to ramping marginal price ratio in h$^{-1}$.} 
\nomenclature{$Q_0$}{Initial load in MW.}
\nomenclature{$Q_z$}{Must-take generation in MW.}
\nomenclature{$\dot Q_0$}{Initial ramping in MW/h.}
\nomenclature{$\dot Q_T$}{Terminal ramping in MW/h.}
\nomenclature{$E_\Delta$}{Energy demand parameter in MWh.}
\nomenclature{$Q_\Delta$}{Power demand parameter in MW.}
\nomenclature{$A$}{A cost parameter (unit varies according to context).}
\nomenclature{$B$}{A cost parameter (unit varies according to context).}
\nomenclature{$C$}{A cost parameter (unit varies according to context).}
\nomenclature{$D$}{A cost parameter (unit varies according to context).}

\section{Introduction}

The growth of renewable electricity generation resources is driven in part by climate-change mitigation policies that seek to reduce the long-term societal costs of continued dependence on fossil-based electricity generation and meet growing electric system load using lower cost resources.  However, each class of renewable generation comes with one or more disadvantages that can limit the degree to which they may be effectively integrated into bulk system operations.  

Hydro-electric generation has long been employed as a significant renewable electric energy and ramping resource. But climate change may jeopardize the magnitude and certainty with which the existing assets can meet demand \cite{markoff2008,chandell2011}. Concerns about population displacement, habitat loss and fishery sustainability often limit the growth of new hydro-electric generation assets, placing additional constraints on new ramping response resources, such as requiring the use of additional reserves and ramping resources.  Shifts in both load and hydro-electric generation potentially increase uncertainty in long term planning and further enhance the need for technological configurations that support  operational flexibility \cite{parkinson2015}. 

Wind power has seen rapid growth, but concern about system reliability has limited the amount of wind generation that can be supported without additional planning and operational measures, such as committing more carbon-intensive firming resources \cite{ortega2009}. Solar resources are also becoming increasingly available but the intermittency challenges are similar to those of wind. In addition, residential rooftop solar resources are challenging the classical utility revenue model \cite{blackburn2014}, can cause voltage control issues in distribution systems \cite{alam2013}, and in extreme cases can result in overgeneration \cite{denholm2015overgeneration}. Taken together these considerations have given rise to questions about the reliable, robust control and optimal operation of an increasingly complex bulk electricity system \cite{dobson2007}.

The conventional utility approach to addressing renewable generation variability is to allocate additional firm generation resources to replace all potentially non-firm renewables resources.  These firm resources are typically fast-responding thermal fossil resources or hydro resources when and where available.  For new renewable resources the impact of this approach is quantified as an intermittency factor, which discounts for instance the contribution of wind in addition to its capacity factor and limits the degree to which renewables can contribute to meeting peak demand \cite{boyle2012}.  However, the intermittency factor does not account for the ramping requirements created by potentially fast-changing renewable resources \cite{makarov2009}.  The need for fast-ramping resources discourages the dispatch of high-efficiency fossil and nuclear generation assets and can encourage reliance on low-efficiency fossil-fuel resources for regulation services and reserves \cite{nyberg2013thermal}.

One solution to overcoming the renewable generation variability at the bulk electric level is to tie together a number of electric control areas into a super-grid so that they can share generation and reserve units through optimal scheduling of system interties \cite{behboodi2017interconnection}. In an interconnected system, the combined power fluctuations are smaller than the sum of the variations in individual control areas. Furthermore, fast-acting energy storage systems and demand response programs can provide required ancillary services such as real-time power balancing \cite{behboodi2017transactive} and frequency regulation \cite{chassin2017H2optimal} if they are equipped with suitable control mechanisms. A competitive market framework in which energy resources participate to sell and buy ancillary service products can accelerate the transition to a high-renewable scenario by supporting the long-term economic sustainability of flexible resources.

Concerns about the financial sustainability of utilities under high level of renewables are also beginning to arise. The question is particularly challenging when one seeks solutions that explicitly maximize social welfare rather than simply minimizing production cost \cite{stern2006objective}.  The growth of low-marginal cost renewable resources can lead one to expect utility revenues to decline to the point where they can no longer recover their long term average costs. But this conclusion may be erroneous if one fails to consider both the impact of demand own-price elasticity, as well as the impact of load control automation on substitution elasticity.  The latter type of demand response can significantly increase the total ramping resource on peak and decrease ramping resource scarcity.  One option for replacing energy resource scarcity rent is increasing fixed payments. But this may lead to economic inefficiencies as well as an unraveling of the market-based mechanisms built so far. Another option is to enable payments based on ramping resource scarcity rent through existing markets for ancillary services.  At the present time, the majority of resources continue to be dispatched based on the energy marginal cost merit order. But it is not unreasonable to consider how one might operate a system in which the energy price is near zero and resources are dispatched instead according the ramping cost merit order.

In the presence of high levels of variable generation, the scheduling problem is a co-optimization for allocating energy and ramping resources \cite{tan2006cooptimization}. Under existing energy deregulation policies, there is usually a market in which energy producers compete to sell energy, and a separate market in which they compete to sell power ramping resources for flexibility. Producers get paid for their energy deliveries in the energy market and for power ramping flexibility in the flexibility market. But today's dual-pricing mechanism is dominated by the energy markets, which drives generation resources to secure revenue primarily in the energy market, and only deliver residual ramping resources in the flexibility market. 
Meanwhile poor access to energy markets leads loads and storage to seek participation primarily in the flexibility market while only revealing their elasticities to the energy market. This relegates loads and storage to only a marginal role in the overall operation of the system, which is the motivation for seeking policy solutions to improving their access to wholesale energy markets, such as FERC Orders 745 and 755.

This paper proposes a system-wide optimal resource dispatch strategy that enables a shift from primarily energy cost-based approach to primarily ramping cost-based one.  This optimal dispatch answers the question of what power schedule to follow during each hour as a function of the marginal prices of energy, power and ramping over the hour\footnote{We define the marginal price of a product or service as the change in its price when the quantity produced or delivered is increased by one unit.}. The main contributions of this paper are (1) the derivation of the formal method to compute the optimal sub-hourly power trajectory for a system when the cost of energy and ramping are both of the same order, (2) the development of an optimal resource allocation strategy based on this optimal trajectory, and (3) a simulation method to evaluate the cost savings of choosing the optimal trajectory over the conventional sub-hourly dispatch used in today's system operation.

In Section~\ref{sec:methodology} we develop the optimal control function in both time and frequency domains. In the case of the frequency domain optimal control function the solution is presented as a continuous function. A discrete-time solution suitable for periodic feedback control systems is presented in Section~\ref{sec:design}.  
In Section~\ref{sec:performance} we examine the performance of this optimal dispatch solution in terms of varying prices for a given ``typical'' hour and in Section~\ref{sec:weccstudy}, we analyze the cost savings in an interconnection that models the Western Electric Coordinating Council (WECC) system for the year 2024 under both low (13\%) and high (50\%) renewable generation scenarios.  Finally, in Section~\ref{sec:discussion} we discuss some of the consequences that appear to arise from this new paradigm and our perspectives on possible future research on this topic.

\section{Methodology} \label{sec:methodology}

Consider a utility's cost minimization problem over a time interval $T$. The utility's customers purchase their net energy use $E(T)$ at a pre-determined retail price. So in today's systems, profit maximization and cost minimization are essentially the same problem. For each hour the utility pays for energy delivered at a real-time locationally-dependent wholesale price that is also dependent on demand under typical deregulated nodal pricing markets. The utility's scheduled energy use is forecast for each hour based on their customers' expected net energy use, which is then used to compute the utility's net load over that hour. We assume that over any interval $T$ the utility may incur additional costs for any deviation in actual net load from the scheduled load.   

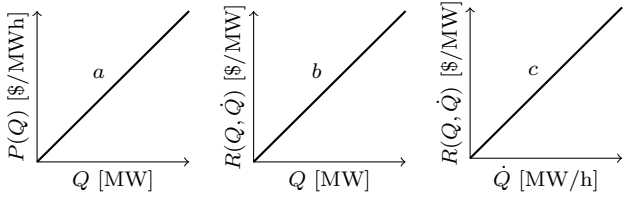
\begin{figure}[!t]
	\centering
	{\footnotesize 
	\begin{tikzpicture}[scale=2]
		\draw [<->] (0,1)--(0,0)--(1,0);
		\node [below] at (0.5,0) {$Q$ [MW]};
		\node [above,rotate=90] at (0,0.5) {$P(Q)$ [\$/MWh]};
		\draw [thick] (0,0)--(1,1);
		\node [above left] at (0.5,0.5) {$a$};
	\end{tikzpicture}
	~
	\begin{tikzpicture}[scale=2]
		\draw [<->] (0,1)--(0,0)--(1,0);
		\node [below] at (0.5,0) {$Q$ [MW]};
		\node [above,rotate=90] at (0,0.5) {$R(Q,\dot Q)$ [\$/MW]};
		\draw [thick] (0,0)--(1,1);
		\node [above left] at (0.5,0.5) {$b$};
	\end{tikzpicture}
	~
	\begin{tikzpicture}[scale=2]
		\draw [<->] (0,1)--(0,0)--(1,0);
		\node [below] at (0.5,0) {$\dot Q$ [MW/h]};
		\node [above,rotate=90] at (0,0.5) {$R(Q,\dot Q)$ [\$/MW]};
		\draw [thick] (0,0)--(1,1);
		\node [above left] at (0.5,0.5) {$c$};
	\end{tikzpicture}
	}
	\caption{Power (left) and ramp (center and right) price functions.}
	\label{fig:price_functions}
\end{figure}

The price function at the operating point is split up into the marginal price of energy $a = \pd{P}{Q}$ (measured in \$/MW$^2\cdot$h), the marginal price of power $b = \pd{R}{Q}$ (measured in \$/MW$^2$), and the marginal price of ramping $c = \pd{R}{\dot Q}$ (measured in \$$\cdot$h/MW$^2$).  In order to reflect resource scarcity all cost functions are assumed to be quadratic so that the price function for each is linear as shown in \reffig{price_functions}. 
The marginal prices $a$ and $b$ determine prices as a function of the power demand $Q$, and the marginal price $c$ determines prices based on the ramp rates $\dot Q$. The cost parameters arise from the schedule and may vary from hour to hour, but do not change within any given hour. Any of the marginal prices may be zero or positive depending on the market design and prevailing conditions in the system.  For the purposes of this paper, we will assume that they cannot be negative.

Over the time interval $T$ the total cost of both the power trajectory $Q(t)$ and the ramping trajectory $\dot Q(t)$ given the power price $P(t) = a Q(t)$ and ramp price $R(t) = b Q(t) + c \dot Q(t)$, respectively, is given by
\eqn {cost} {
	C(T) & = \int_0^T P[Q(t)] Q(t) + R[Q(t),\dot Q(t)] \dot Q(t) dt.
}
Given the dispatch from $Q(0)$ to $Q(T)$ and the scheduled energy use $E(T) = \int_0^T Q(t) dt$ we augment the cost function with the Lagrange multiplier $\lambda$ so that we have
\eq {
	\int_0^T a (Q-Q_z) Q + b (Q-Q_z) | \dot Q | + c \dot Q^2 + \lambda Q ~ dt 
\\
	= \int_0^T G(t,Q,\dot Q) dt,
}
where the $|\dot Q|$ represents the magnitude of the ramp rate $\dot Q$, and $Q_Z$ is the amount of must-take generation having zero or effectively zero marginal energy cost. Then the optimal dispatch trajectory $Q(t)$ is the critical function obtained by solving the Euler-Lagrange equation
\eq {
	\pd{G}{Q} - \dd{}{t} \pd{G}{\dot Q} = 0.
}
From this we form a second-order ordinary differential equation describing the critical load trajectory
\eq {
	\ddot Q - \frac{a}{c} Q = \frac{\mu}{2c}.
}
where $\mu = \lambda - a Q_Z$. Using the Laplace transform we find the critical system response in $s$-domain is
\eqn {solution} {
	\hat Q(s) 
	&= \frac{Q_0 s^2 + \dot Q_0 s + \frac{\mu}{2c}}{s(s^2-\omega^2)},
} 
where $\omega^2 = \tfrac{a}{c}$. The general time-domain solution for the critical function over the interval $0 \leq t<T$ is
\eqn {solution1} {
	Q(t) = \left( Q_0 + \frac{\mu}{2a} \right) \cosh{\omega t} + \frac{ \dot Q_0}{\omega} \sinh{\omega t} - \frac{\mu}{2a},
} 
where $Q_0$ and $\dot Q_0$ are initial power and ramp values.

We can determine whether this solution is an extremum by computing the second variation
\eq {
	\pdn2{C}{Q}(T) & = \int_0^T [\alpha (v)^2+2\beta(vv')+\gamma(v')^2]dt 
\\
	& = \int_0^T H(t) dt,
}
with $H(t)>0$ for all $v\ne0$ subject to $v(0)=0=v(T)$.
We then have
\eq {
		\alpha=\pdn2{G}{Q}=2a, 
	\quad
		\beta=\pd{^2G}{Q\partial\dot Q}=b,
	\quad
		\gamma=\pdn2{G}{\dot Q}=2c.
}
Thus for all $a,b,c>0$, $H(t)>0$ and $Q(t)$ is a minimizer. Since the only physical meaningful non-zero values of $a$ and $c$ are positive, this is satisfactory. We will examine cases when $a$ and $c$ are zero separately. Note that when $\dot Q<0$, we have $b<0$, so that the sign of $b$ does not affect the general solution.

Given the constraints $\int_0^T Q(t) dt = E_T$ and $Q(T) = Q_T$, which come from the hour-ahead schedule, we obtain the solution for $\mu$ and $\dot Q_0$ for the case where $a,c>0$:
\eqn {parameters} {
	\Matrix{\mu\\\dot Q_0} = \Matrix{A&B\\C&D}^{-1} \Matrix{E_\Delta\\Q_\Delta},
}
where
\eq {
		A &= \frac{\sinh{\omega T}-\omega T}{2a\omega}
	&	B &= \frac{\cosh{\omega T}-1}{\omega^2}
\\		C &= \frac{\cosh{\omega T}-1}{2a}
	&	D &= \frac{\sinh{\omega T}}{\omega}
\\		E_\Delta &= E_T - \frac{\sinh{\omega T}}{\omega} Q_0
	&	Q_\Delta &= Q_T - Q_0 \cosh{\omega T}.
}
 
When $a=0$, the cost of energy is zero and only the ramping cost is considered. Then the time-domain solution is
\eqn {solution2} {
	Q(t) = \frac{\mu}{4c}t^2 + \dot Q_0 t + Q_0,
}
with 
\eq {
		A &= \frac{T^3}{12c}
	&	B &= \frac{T^2}{2}
\\		C &= \frac{T^2}{4c}
	&	D &= T
\\		E_\Delta &= E_T - Q_0 T 
	&	Q_\Delta &= Q_T - Q_0,
}
which gives the critical response in $s$-domain
\eq {
	\hat Q(s) = \frac{\mu}{4cs^3} + \frac{\dot Q_0}{s^2} + \frac{Q_0}{s}.
}

When $c=0$, there is no scarcity for ramping so that the ramping price is based only on the marginal energy cost of additional units that are dispatched. Then we have the time-domain solution
\eqn {solution3} {
	Q(t) = -\frac{\mu}{2a},
}
with 
\eq {
	\mu = -\frac{2aE_T}{T}.
}
This gives the critical response is $s$-domain
\eq {
	\hat Q(s) = -\frac{\mu}{2as},
}
and the initial and final ramps from $Q(0)$ to $-\frac{\mu}{2a}$ and from $-\frac{\mu}{2a}$ to $Q(T)$ are limited by the ramping limits of the responding units. 

\section{Optimal Dispatch Controller}
\label{sec:design}

The partial fraction expansion of Eq.~\ref{eq:solution} is
\eqn {controller} {
	\frac{K_1}{s+\omega} + \frac{K_2}{s} + \frac{K_3}{s-\omega},
}
where $K_1=\frac{Q_0}{2}-\frac{\dot Q_0}{2\omega}+\frac{\mu}{4a}$, $K_2=-\frac{\mu}{2a}$, and $K_3=\frac{Q_0}{2}+\frac{\dot Q_0}{2\omega}+\frac{\mu}{4a}$, with the values of the parameters are computed from Eq.~\ref{eq:parameters}. 

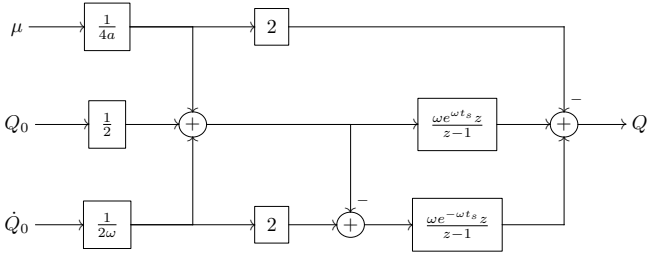
\begin{figure}[!t]
	\centerline { \scalebox{0.75}{
		\xymatrix {  
			\mu \ar[r]
		&	*++[F]{\frac{1}{4a}} \ar[rr] \ar`r/0pt[rd][rd]
		&	
		&	*++[F]{2} \ar`r[rrrd][rrrd]^>>>-
		&
		&	
		&
		\\	Q_0 \ar[r]
		&	*++[F]{\frac{1}{2}} \ar[r]
		&	*+o[F]{+} \ar[rrr] \ar`r[rrd][rrd]^>>>-
		&
		&
		&	*++[F] {\frac{\omega e^{\omega t_s} z}{z-1}}
				\ar[r]
		&	*+o[F]{+} \ar[r]
		&	Q
		\\	\dot Q_0 \ar[r]
		&	*++[F]{\frac{1}{2\omega}} \ar[rr] \ar`r[ru][ru]
		&
		&	*++[F]{2} \ar[r]
		&	*+o[F]{+} \ar[r]
		&	*++[F] {\frac{\omega e^{-\omega t_s} z}{z-1}}
				\ar`r[ru][ru]
		&	
		}
	}}
	\caption{Optimal dispatch controller diagram with discrete update time $t_s$.}
	\label{fig:controller}
\end{figure}

The initial response of the optimal controller is dominated by the forward-time solution
\eq {
	K_1 ~ \exp { - \omega t } = \mathcal{L}^{-1} \left[ \frac{\frac{Q_0}{2}-\frac{\dot Q_0}{2\omega}+\frac{\mu}{4a}}{s+\omega}\right](s),
}
which handles the transition from the initial system load $Q_0$ to the scheduled load $Q_E = - \tfrac{\mu}{2 a s}$. 
The central response is dominated by the scheduled load solution
\eq {
	K_2 = \mathcal{L}^{-1} \left[ - \frac{\mu}{2 a s} \right](s).
}
Finally, the terminal response is dominated by the reverse-time solution
\eq {
	K_3 ~ \exp { \omega t } = \mathcal{L}^{-1} \left[ \frac{\frac{Q_0}{2}+\frac{\dot Q_0}{2\omega}+\frac{\mu}{4a}}{s-\omega}\right](s),
}
which handles the transition from the scheduled load to the terminal load $Q_T$.
A discrete-time controller that implements the solution of Eq.~\ref{eq:controller} is shown in Figure~\ref{fig:controller}. The controller implements the three main components to the optimal solution with step inputs $\mu$, $Q_0$, and $\dot Q_0$. Note that the marginal prices $a$, $b$ and $c$ for the entire hour are constants in the controller blocks, which makes the controller design linear time-invariant within each hour, but time-variant over multiple hours. The discrete-time solution is then
\eq {
	Q^*(k) = \left\{ \begin{array}{ll}
		K_1 \tau^{k} + K_2 + K_3 \tau^{-k} &: a>0,c>0
	\\
		\frac{\mu}{4c} t_s^2 k^2
		+ \dot Q_0 t_s k + Q_0 &: a=0, c>0
	\\
		-\frac{\mu}{2a} &: a>0, c=0
	\end{array} \right.
}
where $\tau=\exp{\omega t_s}$. 

\begin{figure}[!t]
	\centering
	\includegraphics[width=\columnwidth,clip,viewport=10 00 400 310]{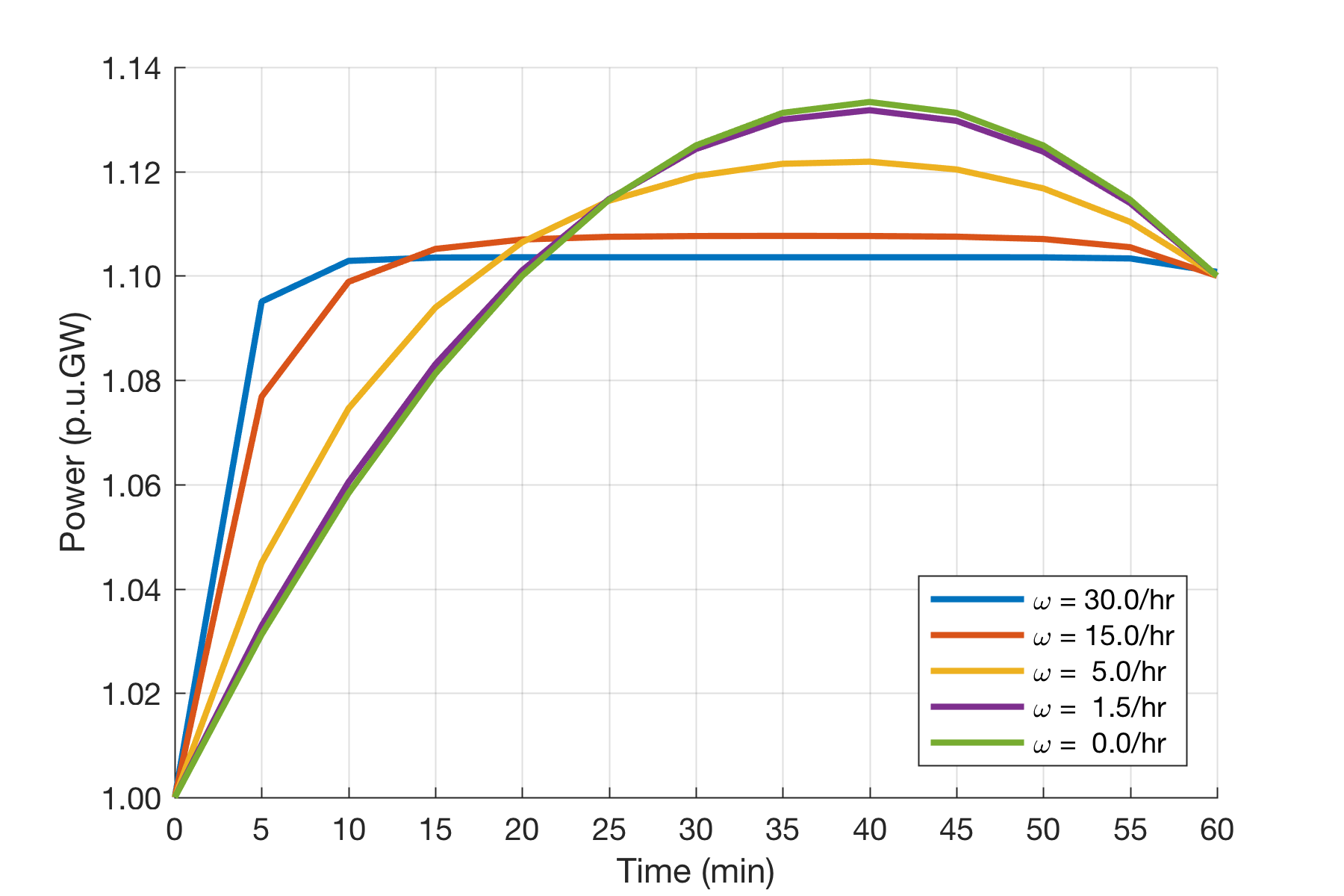}
	\caption{Optimal discrete time control for various values of $\omega$ at $t_s=5$ minutes.} 
	\label{fig:optimal_ramp}
\end{figure}

The discrete-time dispatch control is illustrated in Figure~\ref{fig:optimal_ramp} for various values of $\omega=\sqrt{a/c}$.
When the value of $\omega$ is large, the optimal dispatch is dominated by the energy cost and the cost of high ramp rates is negligible compared to the energy cost. 
The result is a dispatch that moves as quickly as possible to scheduled load $Q_E$. In the limit of zero ramping cost, the optimal response is a step function\footnote{Step responses are only possible by generation or load tripping, which is not considered as part of the conventional control strategy.}. As the cost of ramping increases relative to the energy cost, the optimal dispatch begins to reduce the ramp rate while still following a trajectory that satisfies the hourly energy delivery requirement.  In the limit of zero energy cost, the optimal dispatch trajectory is parabolic.

\section{Performance Evaluation} \label{sec:performance}

In this section we develop the cost performance metric of the optimal dispatch control design.  The optimal dispatch cost function is found by evaluating Equation~\ref{eq:cost} using Equations~\ref{eq:solution1}, \ref{eq:solution2} and \ref{eq:solution3}.  Thus when $a,b,c>0$ we have\footnote{Note that if the ramp rate $\dot Q$ changes sign at the time $t_c=\frac{1}{\omega} \tanh^{-1} (-\frac{B}{A})$ and $0<t_c<T$, then we must divide the cost integral into two parts to account for the absolute value of $\dot Q$ on $b$ terms.} 
\eq {
	C(T) &= \frac{\sinh 2\omega T}{2\omega} \left[ a(A^2+B^2) + bAB\omega \right]
	\\&	+ \frac{\sinh^2{\omega T}}{2\omega} \left[ b(A^2+B^2)\omega + 4aAB \right]
	\\&	+ \frac{\cosh{\omega T}-1}{\omega} [(bA\omega + 2aB)C - (aB+bA\omega) Q_z]
	\\&	+ \frac{\sinh{\omega T}}{\omega} [(bB\omega + 2aA)C -(aA+ bB\omega) Q_z]
	\\&	+ \left[aC^2 - aC Q_{z} \right]T.
}
where $A=Q_0+\mu/2a$, $B=\dot Q_0/\omega$ and $C=-\mu/2a$.
For the case when $a=0$ we have 
\eq {
	C(T) &= 
	\tfrac12 b A^2 T^4
	 + [bB+\tfrac43 cA]A T^3
	\\& + [\tfrac12 b(B^2+2AC)+2cAB -bAQ_z] T^2
	\\& + [(bC+cB)B -bBQ_z] T
	,
}
where $A=\mu/4c$, $B=\dot Q_0$, and $C=Q_0$.
When $c=0$ we have
\eq {
	C(T) &= a E_T \left( \frac{E_T}{T} - Q_z \right).
}

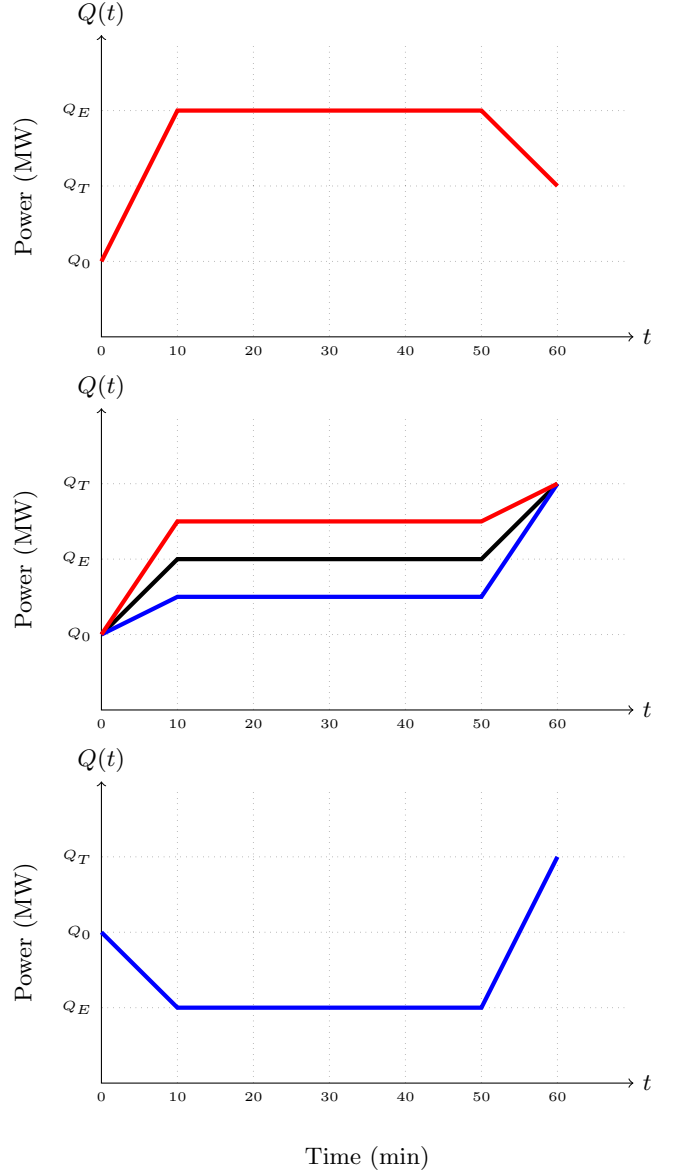
\begin{figure}[!t]
	\centerline {
		\begin{tikzpicture}
			\draw [<->] (0,4)--(0,0)--(7,0);
			\draw [ultra thin,dotted] (0.01,0.01) grid (6.9,3.9);
			\node [right] at (7,0) {\small $t$};
			\foreach \n in {0,10,...,60}
			{
				\node [below] at (\n/10,0) {\tiny $\n$};
			}	
			\node [above] at (0,4) {\small $Q(t)$};
			\node [left] at (0,1) {\tiny $Q_0$};
			\node [left] at (0,2) {\tiny $Q_T$};
			\node [left] at (0,3) {\tiny $Q_E$};
			\node [rotate=90] at (-1,2) {\small Power (MW)};
			\draw [ultra thick,red] (0,1)--(1,3)--(5,3)--(6,2); 
		\end{tikzpicture}
	}
	\centerline {
		\begin{tikzpicture}
			\draw [<->] (0,4)--(0,0)--(7,0);
			\draw [ultra thin,dotted] (0.01,0.01) grid (6.9,3.9);
			\node [right] at (7,0) {\small $t$};
			\foreach \n in {0,10,...,60}
			{
				\node [below] at (\n/10,0) {\tiny $\n$};
			}	
			\node [above] at (0,4) {\small $Q(t)$};
			\node [left] at (0,1) {\tiny $Q_0$};
			\node [left] at (0,2) {\tiny $Q_E$};
			\node [left] at (0,3) {\tiny $Q_T$};
			\node [rotate=90] at (-1,2) {\small Power (MW)};
			\draw [ultra thick] (0,1)--(1,2)--(5,2)--(6,3); 
			\draw [ultra thick,blue] (0,1)--(1,1.5)--(5,1.5)--(6,3); 
			\draw [ultra thick,red] (0,1)--(1,2.5)--(5,2.5)--(6,3); 
		\end{tikzpicture}
	}
	\centerline {
		\begin{tikzpicture}
			\draw [<->] (0,4)--(0,0)--(7,0);
			\draw [ultra thin,dotted] (0.01,0.01) grid (6.9,3.9);
			\node [right] at (7,0) {\small $t$};
			\node at (3.5,-1) {\small Time (min)};
			\foreach \n in {0,10,...,60}
			{
				\node [below] at (\n/10,0) {\tiny $\n$};
			}	
			\node [above] at (0,4) {\small $Q(t)$};
			\node [left] at (0,2) {\tiny $Q_0$};
			\node [left] at (0,1) {\tiny $Q_E$};
			\node [left] at (0,3) {\tiny $Q_T$};
			\node [rotate=90] at (-1,2) {\small Power (MW)};
			\draw [ultra thick,blue] (0,2)--(1,1)--(5,1)--(6,3);
		\end{tikzpicture}
	}
	\caption{Conventional power dispatch for base case: (top) significant negative schedule error requiring over-production, (center) small negative, zero and positive schedule error requiring over (red), normal (black) and under (blue) production, and (bottom) significant positive schedule error requiring under-production.} 
	\label{fig:basecase}
\end{figure}

We use as the base case a conventional unit dispatch strategy that requires generators ramp to their new operating point during the 20 minutes spanning the top of the hour. Accordingly the generators begin ramping 10 minutes before the hour and end ramping 10 minutes after the hour.  In the aggregate for a given hour this strategy is illustrated in Figure~\ref{fig:basecase} where
\eq {
	Q_E = \frac{6}{5} \left( E_T - \frac{Q_0+Q_T}{12} \right),
}
with the initial and terminal ramp rates
\eq {
	\dot Q_0 = 6(Q_E-Q_0)
	\qquad \mathrm{and} \qquad
	\dot Q_T = 6(Q_T-Q_E).
}
Three cases are shown: overproduction to compensate for a lack of generation in previous hours (top), scheduled production (center), and underproduction to compensate for extra generation in previous hours (bottom).

The cost of the base case is then
\eq {
	C_{base}(T) &= \tfrac{a T}{18} (Q_T^2 + Q_T Q_E + 14 Q_E^2 + Q_E Q_0 + Q_0^2) 
	\\ & - \tfrac{a T}{12} (Q_T + 10 Q_E + Q_0) Q_z 
	\\ & + |\tfrac{b}{2} (Q_E - Q_0)|(Q_E+Q_0-2Q_z) 
	\\ & + |\tfrac{b}{2} (Q_T - Q_E)|(Q_T+Q_E-2Q_z)   
	\\ & + \tfrac{6 c}{T} (Q_T^2 - 2 Q_T Q_E + 2 Q_E^2 - 2 Q_0 Q_E + Q_0^2).	
}

The zero-order hold ramp discrete form of Equation~\ref{eq:cost} gives us the cost of operating with a discrete control time-step $t_s$, i.e.,
\eq {
	C^*(T) & = \sum_{k=0}^{T/t_s} \left( P^*[Q^*(k)] Q^*(k) + R^*[Q^*(k),\dot Q^*(k)] \dot Q^*(k) \right) t_s
\\ 
	& = \sum_{k=0}^{T/t_s}
	   \tfrac{at_s}{4} \left[Q^*(k)^2 + 2Q^*(k)\dot Q^*(k) \right.
\\
	&	\qquad \qquad + \left. \dot Q^*(k)^2 - 2Q_z [ Q^*(k) + \dot Q^*(k) ]\right] 
\\ 
	& \quad + \tfrac{1}{2} \left[ \left| b (\dot Q^*(k)- Q^*(k)) \right| (\dot Q^*(k) + Q^*(k) - 2Q_z \right] 
\\ 
	& \quad + \tfrac{c}{t_s} \left[ Q^*(k)^2 - 2Q^*(k) \dot Q^*(k) + \dot Q^*(k)^2 \right] 
}
where $Q^*(k)=Q(k t_s)$ and $\dot Q^*(k) = Q^*(k+1)$. We evaluate the performance of the control strategy for different control update rates $t_s$ using two future scenarios, one for low renewables where $\omega > 1$, and one for high renewables where $\omega < 1$ for both unconstrained and constrained transmission operating conditions. 

\section{Case Study: WECC 2024}
\label{sec:weccstudy}

In this section we examine the cost savings associated with using the optimal control solution on the WECC 2024 base case model introduced in \cite{behboodi2017interconnection}.  The WECC 2024 model is a 20-area base case used by WECC for planning studies.  The 20-area model combines a number of smaller control areas based on the anticipated intertie transfer limits reported in the WECC 2024 common case \cite{wecc2024commoncase}.  In this model constraints within control areas are ignored, while internal losses are approximated.  The peak load, annual energy production and demand consumption are forecast, including intermittent wind, solar, and run-of-river hydro for the entire year.

The model also includes a hypothetical market for each consolidated control area, with a flat zero-cost supply curve for all renewable and must-take generation resources and a constant positive supply curve slope for all dispatchable units.  The hourly generation of intermittent resources is provided by the base case model and incorporated into the supply curve so that there is effectively no marginal cost of production for renewable energy and must take generation.  All generating units are paid the hourly clearing price, and when the marginal energy price in a control area is zero then renewable generation may be curtailed.  As a result, under the high renewable scenario, zero energy prices are commonplace and renewable generation is curtailed more frequently. Demand response is similarly considered for each control area and the output of this scheduling model provides the hourly nodal prices required to satisfy the transmission constraints, if any. 

The low renewables case is the WECC forecast for the year 2024, which correspond to 29.5 GW (16.1\%) of renewable capacity and 140.8 TWh (13.4\%) of annual renewable generation.  The high renewables case is given as 400\% of capacity of the WECC forecast for the year 2024, which corresponds to 117.8 GW (63.5\%) and 523.9 TWh (49.6\%), respectively. The blended energy price of operations is \$130.6/MWh and \$50.2/MWh for the low and high renewables cases, respectively.

The ramping price was not considered in the WECC 2024 base case model.  For this study we have assumed that the ramping energy cost is based on the marginal energy cost for the dispatchable generation and the demand response, as well as the cost of changing the dispatchable generation output, as shown in Table~\ref{tab:prices}. In both cases, the marginal price of power $b$ is the average marginal price of energy $a$ over the hour. In the low renewables case the marginal price of ramping $c$ is the marginal price of power $b$ multiplied by 12 seconds. In the high renewables case, $c$ is the marginal price of power $b$ multiplied by 49 hours. The value of $\omega$ is approximately 121 times greater in the low renewable case than it is in the high renewable case. Note that $a$ is zero when renewables are curtailed while $b$ is assumed to also be zero because curtailed renewables and demand response are presumed to be dispatchable. 

The values of the ramping response constant $c$ were also selected such that the overall cost of operating the system remains more or less constant when going from the low to high renewables scenarios under the base case. This allows us to evaluate the impact of the optimal control strategy without involving the question of revenue adequacy under the high renewables scenario. Given that there are few markets from which to determine these values, we must be satisfied with this assumption for now.

The statistical nature of the intermittency and load forecast errors and their connection to load following and regulation was studied at length in \cite{diao2012planning}. The authors showed that consolidated control of WECC could yield both cost savings and performance improvements. In particular, the study showed that with high accuracy control 1\% standard deviation in load forecast was expected, with 0\% real-time mean error at 0.15\% standard deviation at peak load. However, for the purposes of a preliminary study like the one presented in this paper, we will consider the scheduling error to be Gaussian with a mean error of 0 MW and a standard deviation of 100~MW. We believe that energy and flexibility markets should be efficient enough to remove all systematic error from the price signals leaving only the random noise that is satisfactorily modeled by Gaussian noise.

\begin{figure}[!t]
\includegraphics[width=\columnwidth,clip,viewport=40 10 530 410]{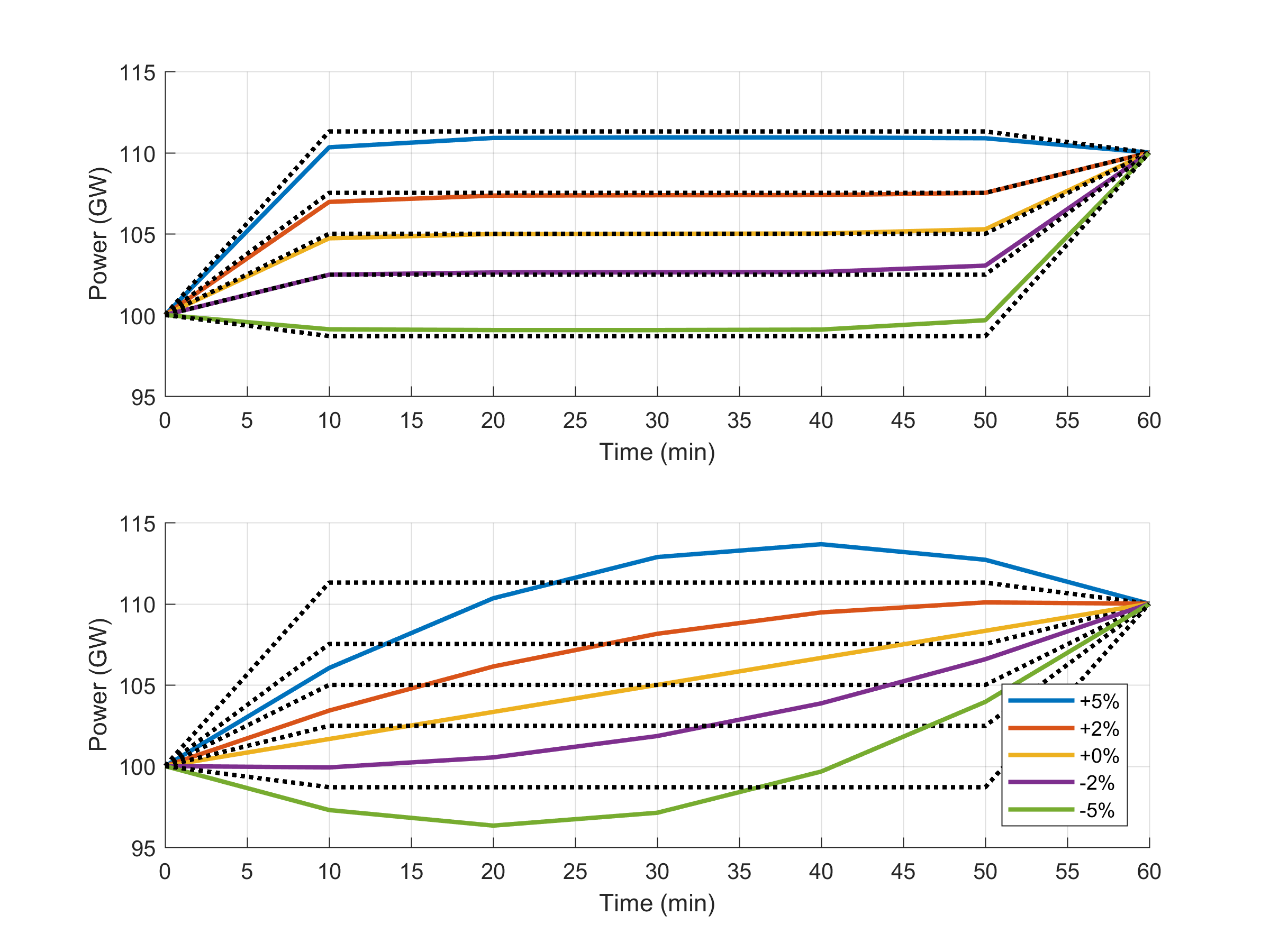}
\caption{Single hour optimal dispatch for low and high renewables with a ramp from 100~GW to 110~GW using a 10-minute discrete-time dispatch control rate, with hourly energy schedule correction errors varying from $-5$\% to $+5$\%.} 
\label{fig:result}
\end{figure}

The comparison of the conventional and optimal dispatch for a typical case is shown in Figure~\ref{fig:result}.  The conventional control strategy is shown in dotted lines, with the 10 minute optimal-dispatch trajectory shown as solid lines. Note that the ramp rate is constant between discrete control updates. The evaluation is completed with the marginal prices and marginal costs at 100~GW, as shown in Table~\ref{tab:prices}. The energy schedule changes according to a varying energy error remaining at the end of the previous dispatch interval.  A $-5$\% error represents an energy deficits of 5~GWh for a 105~GWh schedule, while a $+5$\% error represents an energy surplus of 5~GWh.

\begin{table}[!t]
	\centering
	\caption{Marginal prices and marginal costs for 105~GWh schedule at 100~GW initial power and 10 GW/h ramp for cases in Figure~\ref{fig:result}} 
	\label{tab:prices}
	~\\
	\begin{tabular}{c|ccc} 
	 Variable & Base case & Study case & Units \\ \hline \hline 
	 \multicolumn{4}{l}{\textit{Marginal prices}:} \\ \hline 
	 $a$ & $1.27\times10^{-3}$ & $6.34\times10^{-4}$ & \$/MW$^2\cdot$h \\ 
	 $b$ & $1.27\times10^{-3}$ & $6.34\times10^{-4}$ & \$/MW$^2$ \\ 
	 $c$ & $4.23\times10^{-6}$ & $3.09\times10^{-2}$ & \$$\cdot$h/MW$^2$ \\ 
	 \hline 
	 \multicolumn{4}{l}{\textit{Marginal costs}:} \\ \hline 
	 $P$ & 133.09 & 66.55 & \$/MW$\cdot$h \\ 
	 $R$ & 133.13 & 375.19 & \$/MW \\ \hline \hline 
	 $\omega$ & 17.3 & 0.1433 & h$^{-1}$ \\ \hline 
	\end{tabular}
\end{table}

\begin{table}[!t]
	\caption{Single hour cost savings under low and high renewable for a ramp from 100~GW to 110~GW at 5 minute discrete dispatch control update rate, with varying energy error redispatch} 
	\label{tab:results}
	~\\
	\centerline
	{	\scalebox{1.0}{ 
		\begin{tabular}{c|c|c|c} 
		 & \multicolumn{3}{c}{Cost} \\ 
		 Scenario & Base case & Optimal & Savings \\ 
		 Model & (\$B/y) & (\$B/y) & (\$B/y) \\ 
		 \hline 
		 \hline \multicolumn{4}{l}{\textit{Unconstrained:}} \\ \hline 
		 Low & 126.0 & 125.9 & 0.16 (0.1\%) \\ 
		 High & 108.6 & 77.8 & 30.85 (28.4\%) \\ 
		 \hline \multicolumn{4}{l}{\textit{Constrained:}} \\ \hline 
		 Low & 184.4 & 184.1 & 0.26 (0.1\%) \\ 
		 High & 388.3 & 231.2 & 157.12 (40.5\%) \\ 
		\hline \end{tabular}
	}}
\end{table}

The marginal prices in Table~\ref{tab:prices} are chosen to satisfy the following conditions:

\begin{enumerate}

\item The system operating cost is roughly \$100/MWh at a system load of 100~GW.

\item For the low renewables case, the energy cost is roughly 10 times the ramping cost, while for the high renewables case the ramping cost is roughly 10 times the energy cost for the nominal schedule. This was necessary to ensure that costs were the same for both cases.

\item The marginal power price $b$ for both cases is equal to the marginal energy price $a$ of the respective case.

\end{enumerate}

We considered the performance degradation resulting from longer dispatch intervals by evaluating the performance using 5 minute updates, 1 minute updates, and 4 second discrete control timesteps but found no appreciable difference in the economic performance.  The results shown in Table~\ref{tab:results} are shown for the 5 minute dispatch interval. The output of the presented discrete control method is a load profile that does not necessarily lead to the scheduled hourly energy, because the load trajectory over each time intervals (which is linear) is slightly different from the optimal load trajectory (that often has a curvature). One approach to deal with this energy deficiency is to use a higher time resolution, so that the trajectories lay on each other more precisely. Another approach is to adjust the targeted load such that it delivers the scheduled energy over each time interval. In this case, the discrete control load is not necessarily equal to the optimal load.

\begin{figure*}[!t]
	\centering
	\includegraphics[width=\columnwidth,clip,viewport=40 10 560 340]{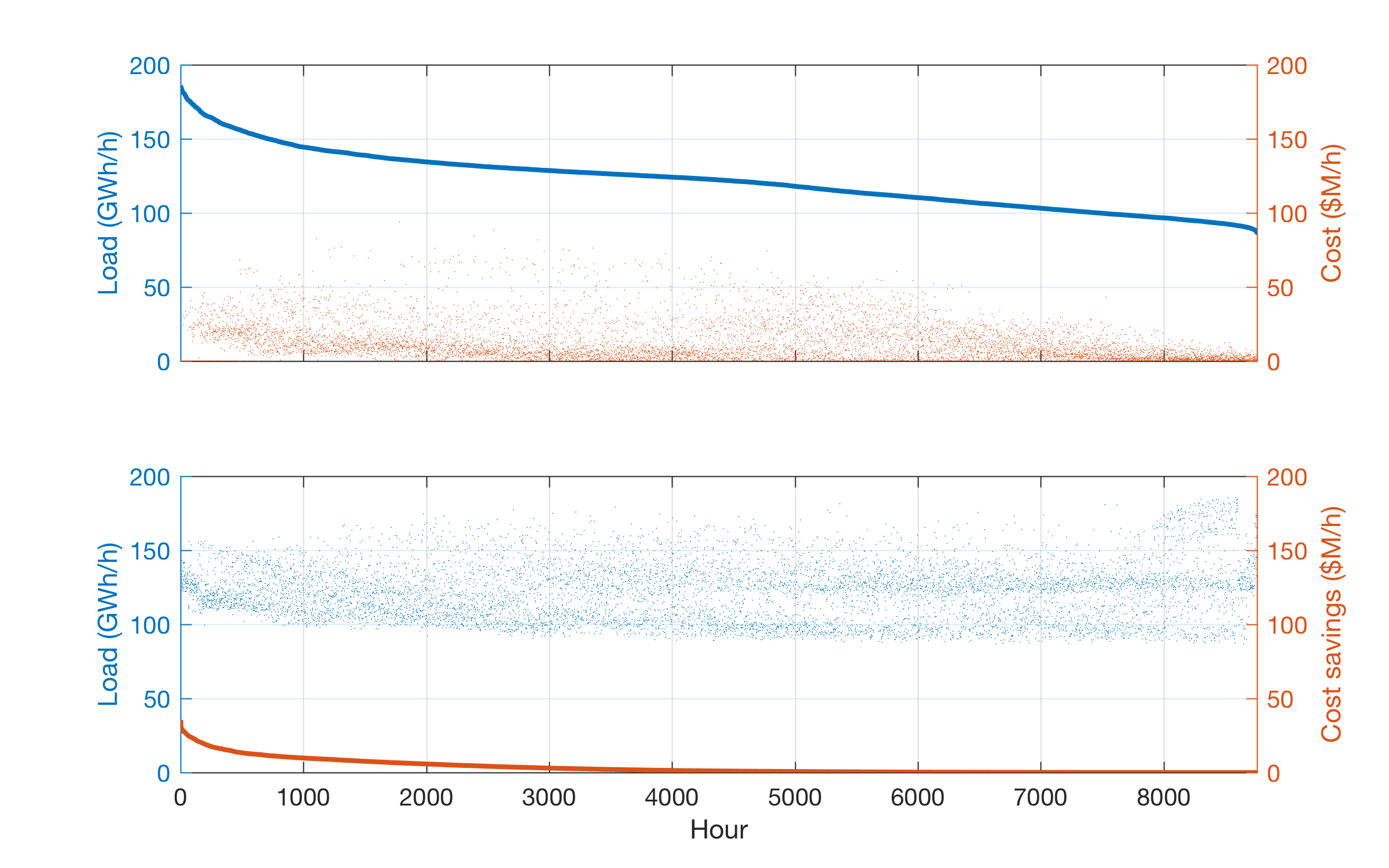}
	\includegraphics[width=\columnwidth,clip,viewport=40 10 560 340]{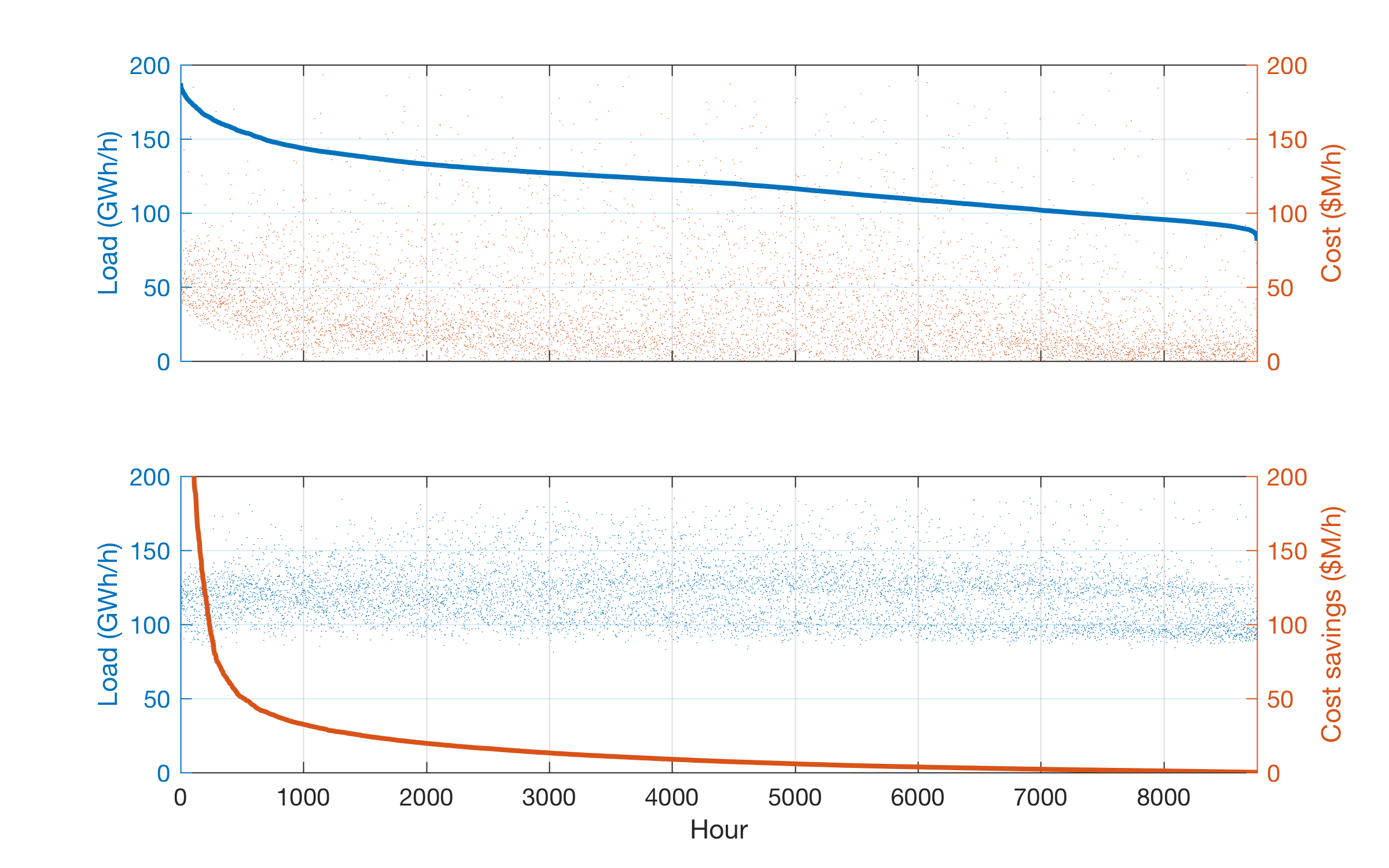}
	\caption{WECC 2024 load duration (top) and optimal dispatch savings duration (bottom) using discrete optimal control at 5-minute dispatch rate for the unconstrained (left) and constrained (right) high renewables scenario. The scatter plots are the corresponding cost (top) and load (bottom) values for the durations curves shown.}
	\label{fig:annual}
\end{figure*}

\begin{table}[!t]
	\centering
	\caption{WECC 2024 cost savings from optimal dispatch under different transmission constraint and renewable scenarios}	
	\label{tab:annual}
	~\\
	\begin{tabular}{c|c|c|c} 
	 & Total & \multicolumn{2}{c}{Price}\\ 
	 Scenario & Energy & Base case & Optimal \\ 
	 Model & (TWh) & (\$/MWh) & (\$/MWh) \\ 
	 \hline 
	 \hline \multicolumn{4}{l}{\textit{Unconstrained:}} \\ \hline 
	 Low & 1054.6 & 119.5 & 119.35 (-0.1\%) \\ 
	 High & 1067.2 & 101.8 & 67.29 (-51.2\%) \\ 
	 \hline \multicolumn{4}{l}{\textit{Constrained:}} \\ \hline 
	 Low & 1054.5 & 174.8 & 174.55 (-0.2\%) \\ 
	 High & 1055.7 & 367.8 & 87.96 (-318.2\%) \\ 
	\hline \end{tabular}
\end{table}

\begin{table}[!t]
	\centering
	\caption{Summary of energy and price impacts of optimal dispatch control for the WECC 2024 base case}
	\label{tab:summary}
	\begin{tabular}{c|c|c|c} 
	 & Total & \multicolumn{2}{c}{Price}\\ 
	 Scenario & Energy & Base case & Optimal \\ 
	 Model & (TWh) & (\$/MWh) & (\$/MWh) \\ 
	 \hline 
	 \hline \multicolumn{4}{l}{\textit{Unconstrained:}} \\ \hline 
	 Low & 1054.6 & 119.5 & 119.35 (-0.1\%) \\ 
	 High & 1067.2 & 101.8 & 67.29 (-51.2\%) \\ 
	 \hline \multicolumn{4}{l}{\textit{Constrained:}} \\ \hline 
	 Low & 1054.5 & 174.8 & 174.55 (-0.2\%) \\ 
	 High & 1055.7 & 367.8 & 87.96 (-318.2\%) \\ 
	\hline \end{tabular}
\end{table}

Generally at low levels of renewables savings are not possible using the optimal control strategy.  The cost savings observed in the extreme low renewables dispatch cases in Table~\ref{tab:results} are due to the fact that discrete dispatch control follows the optimal trajectory sampling every $t_s$ seconds. This dispatch error can result in small over or underproduction depending on the degree of asymmetry in the optimal trajectory.

At higher levels of renewables the savings are potentially more significant. In addition, the savings are maximum when dispatch tracks the original schedule, which suggests that there may be a strong economic incentive to avoid carrying over energy tracking error from one schedule interval to the next.

The interconnetion-wide scheduling solution in \cite{behboodi2017interconnection} includes a 20-area constrained solution. The hourly energy prices for each area are computed considering both supply and demand energy price elasticities. The energy prices are computed for the interconnection-wide surplus-maximizing schedule over the entire year.  The marginal power price is the price of energy for the schedule hour. The marginal price of ramping is 1/300 marginal price of power in the low renewable case, and 49 times the marginal price of power in the high renewable case. The costs, savings and price impact of using this scheduling solution compared to the base case are presented in Tables~\ref{tab:annual} and \ref{tab:summary}. The unconstrained solution is evidently less costly because the combined system-wide fluctuations are smaller than the sum of the individual of the variations in each balancing authority.

The WECC 2024 system-wide load and savings duration curves\footnote{A duration curve shows the number of hours per year that a time-series quantity is above a particular value.  It is obtained by sorting the time-series data in descending order of magnitude and plotting the resulting monotonically descending curve.} are shown in Figure~\ref{fig:annual}. The potential savings are very significant for all scenarios, with the highest savings being found when high levels of renewable resources are available.  The savings when more transmission constraints are active are augmented considerably with respect to unconstrained system conditions.

\section{Discussion} \label{sec:discussion}

The significance of the results shown in Figure~\ref{fig:optimal_ramp} cannot be understated. First we observe that when $a>>c$, the optimal response is very similar to the conventional dispatch strategy, giving us some assurance that today's operations are very nearly optimal. However, when $a<<c$ today's hourly dispatch strategy is not optimal. As the fraction of cost attributed to energy decreases relative to the cost attributed to ramping, we see that $\omega$ decreases and the value of changing the dispatch strategy increases dramatically.  In the limit of a very high renewable scenario the savings achievable using the optimal dispatch strategy can be extremely significant. Failure to adopt an optimal dispatch such as the one proposed could result in major and likely unnecessary costs.  Utilities will inevitably find it necessary to mitigating these costs, either by reducing the amount of renewables, by increasing the revenues from their customers, or by developing some kind of optimal resource allocation strategy such as the one proposed. 

\begin{figure}[!t]
	\includegraphics[width=\columnwidth,clip,viewport=20 0 400 270]{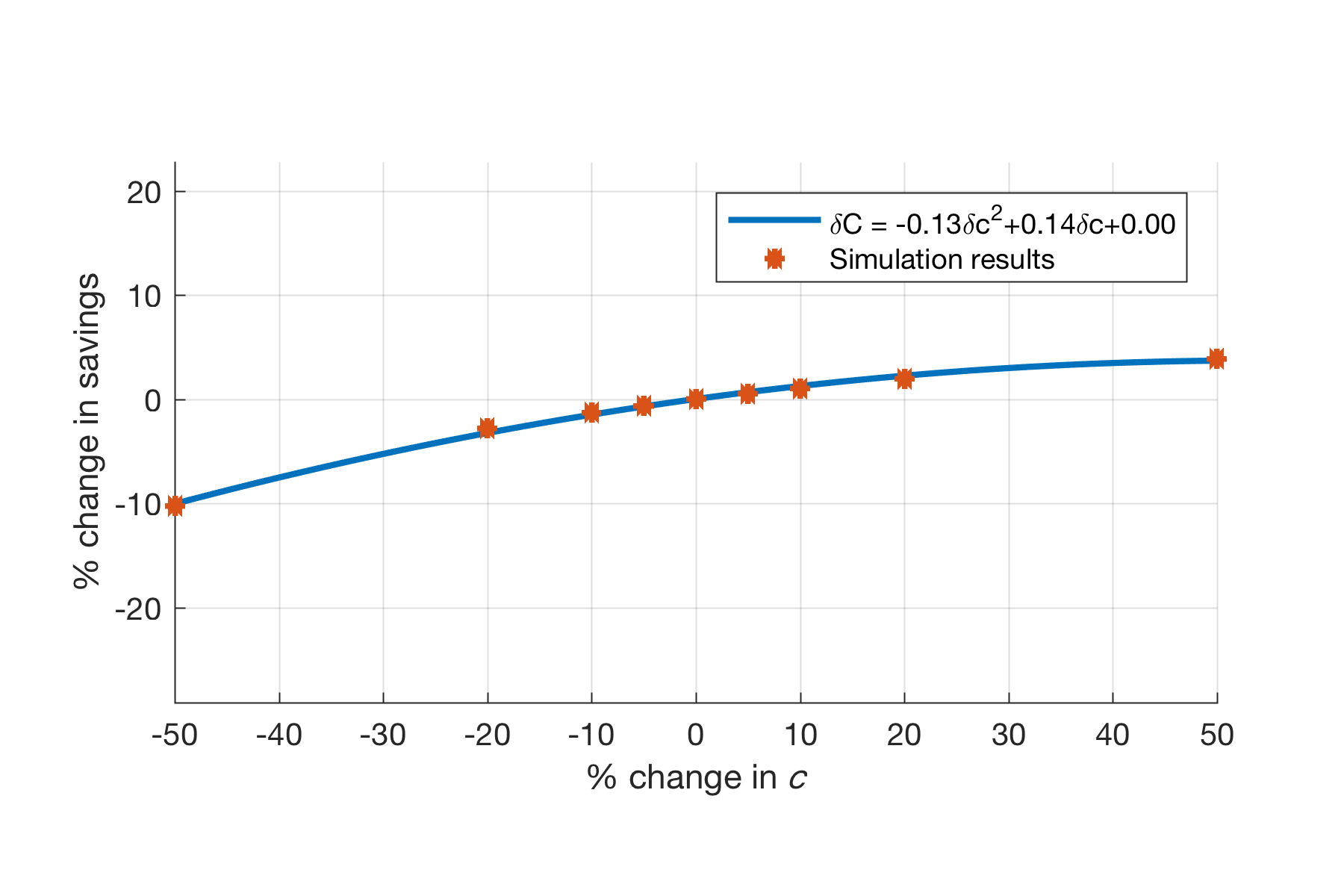}
	\caption{Sensitivity of savings to marginal price of ramping resources.}
	\label{fig:sensitivity}
\end{figure}

A sensitivity analysis of the savings as a function of the marginal price of ramping $c$ shows that the savings are not overly sensitive to changes in our assumption of the cost of ramping scarcity. Figure~\ref{fig:sensitivity} shows that for a 50\% decrease in $c$, we observe a 10.3\% decrease in savings, while a 50\% increase in $c$ results in a 3.9\% increase in savings.  This suggests that the savings from employing the optimal dispatch strategy is quite robust to our uncertainty about the marginal price of ramping resources.

In any financially sustainable future scenario, we must consider how the long-term average costs and fixed costs are recovered under the pricing mechanism.  We have assumed in this study that renewable generation and utilities cannot sustainably continue employing complex power purchasing agreements and subsidies to hedge against energy price volatility. Instead all parties should come to rely on separate real-time pricing mechanisms for energy, power and ramping response of the resources they control.

Shifting revenue from resource allocation mechanisms based primarily on energy resource scarcity to ones based primarily on flexibility resource scarcity can be expected to have a significant impact on the cost of subhourly resource dispatch. The optimal strategy for low renewable conditions very closely matches the strategy employed today when moving hour-to-hour from one scheduled operating point to another. Indeed, the optimal dispatch strategy does not offer any significant cost savings when overall pricing is dominated by energy resource scarcity.

However, as increasing amounts of renewables are introduced, the scarcity rents may shift from energy to flexibility resources. The optimal subhourly dispatch strategy may be expected to change with increasing emphasis on avoiding high ramp rates over sustained periods at the expense of maintaining a constant power level over the hour.

The relationship between existing price signals for various grid services and the three principle price components needed to implement this optimal strategy requires further investigation.  It is evident that the marginal price $a$ represents a linearization of the energy price itself at the current operating point. But it is not clear yet whether and to what degree the marginal prices $b$ and $c$ can be connected to any existing price signals, such as the capacity price or the price of ancillary services like frequency regulation resources, generation reserves, and demand response.  The links do suggest themselves based on both the resource behaviors and physical dimensions of the parameters. However, it is not certain yet whether this will be simply a matter of obtaining a linearization of the services' cost functions at the appropriate operating point.

Additionally, it is instructive to note that the marginal price of redispatched power $b$ is not important to the optimal dispatch strategy, insofar as the parameter does not appear in Eq.~\ref{eq:solution}. This leads one to conclude that to the extent capacity limits do not affect either energy or ramping scarcity rents (or are already captured in them), the marginal cost of additional resource capacity is never considered for optimal subhourly dispatch control. This is consistent with the expectation that sunk costs should not be a factor in the selection of which units to dispatch at what level, at least to the extent that these costs are not entering into the energy or ramping costs.

In the presence of significant renewables, the energy marginal cost does not entirely reflect the grid condition without considering the cost of ramping up and down services.  Therefore, the energy price cannot be solely used as a control signal to the generation and load units to achieve the optimal utilization of resources. In order to quantify the value of ramping product we suggest using a market framework in which flexible generation and load resources compete to sell their ancillary service products at the bulk electric system level. As renewable level rises the energy marginal cost decrease (smaller $a$ value) because renewables are zero-generation cost resources, but the ramping marginal cost increases (larger $c$ value) because the system requires more flexibility to handle the generation variation.  In long run, inflexible units get retired and more flexible units are built to support the renewable integration since flexibility will be a revenue source rather than energy.

The availability of high renewables can lead to situations where low cost energy is being supplied to areas with high cost flexibility resources through constrained interties.  The optimal strategy avoids dispatching these high cost flexilibity resources to the extent possible by reducing the ramping schedule.  The more transmission capacity is available, the lower the overall cost, but we note that even when the system is constrained, the cost of optimally dispatching flexibility resources can be significantly lower under the high renewables case than under a low renewables scenario.

It seems that the use of energy-only market designs run counter to the results of this study.  Flexilibity resource markets may become increasingly important, even in regions that are not dominated by local renewable generation.  This is especially true in cases where adequate transmission capacity is available for renewables in remote regions to displace local dispatchable generation.  This may give rise to a new set of challenges for utility and system operators as they seek a revenue model that not only provides for operating costs, but also maintains the coupling between retail demand response and wholesale supply and retail delivery constraints.  If the cost of the wholesale system becomes increasingly dominated by ramping resource constraints, while retail continues to use energy prices to encourage consumer efficiency, then retail behavior will be not affected as much by short-term wholesale price fluctuations.  This trend runs against the desire for more engaged consumers who can respond to system conditions in real time. Clearly a new utility revenue model is needed if the transformation to a high renewable \textit{modus operandi} is to occur successfully in the coming decades.

\section{Conclusions}

The principal finding of this paper is that the use of an optimal dispatch strategy that considers both the cost of energy and the cost of ramping resources simultaneously leads to significant cost savings in systems with high levels of renewable generation. For the WECC 2024 common case the savings can exceed 25\% of total operating costs in a 50\% renewables scenario. 

As the bulk power interconnection resource mix shifts from primarily dispatchable non-zero marginal cost resources (e.g., natural gas) to primarily non-dispatchable zero marginal cost renewable resources (e.g., run-of-river hydro, wind, solar) we expect a steady shift in bulk system costs from energy resource scarcity rent to ramping resource scarcity rent.  While the total revenue must remain largely the same for financial sustainability, operating strategies must adapt to reflect changes in resource scarcity costs.

\section*{Acknowledgments}

This work was funded in part by the Pacific Institute for Climate Solutions at the University of Victoria and the California Energy Commission under Project EPC-15-047. SLAC National Accelerator Laboratory is operated for the US Department of Energy by Stanford University under Contract No. DE-AC02-76SF00515. The authors also thank Francesco Carducci for his very helpful suggestions during the preparation of this paper.

\footnotesize
\printnomenclature

\section*{References}

\bibliographystyle{unsrt}
\bibliography{references}

\end{document}